\documentclass[11pt]{article}\textwidth 6.5in\textheight 51pc
\usepackage{amssymb}\usepackage[colorlinks]{hyperref}\usepackage{color}
\usepackage{mathrsfs}\usepackage[normalem]{ulem}\usepackage{amsthm}
\usepackage{amsmath}
\usepackage{amssymb}
\usepackage{verbatim}
\usepackage{braket}
\usepackage{setspace}
\usepackage{natbib}
\usepackage{hyperref}

\newcommand{\Tr}{\mathrm{Tr}}

\begin{document}

\newcommand\hreff[1]{\href {http://#1} {\small http://#1}}
\newcommand\trm[1]{{\bf\em #1}}\newcommand\prf{\paragraph{Proof.}}

\newtheorem{thr}{Theorem} 
\newtheorem{lem}{Lemma}
\newtheorem{prp}{Proposition}

\newtheorem{thm}{Theorem} 
\newtheorem{lmm}{Lemma}
\newtheorem{cor}{Corollary}
\newtheorem{con}{Conjecture} 

\newtheorem{blk}{Block}
\newtheorem{dff}{Definition}
\newtheorem{asm}{Assumption}
\newtheorem{rmk}{Remark}
\newtheorem{clm}{Claim}
\newtheorem{example}{Example}

\newcommand\floor[1]{{\lfloor#1\rfloor}}\newcommand\ceil[1]{{\lceil#1\rceil}}

\newcommand{\lea}{<^{+}}
\newcommand{\gea}{>^{+}}
\newcommand{\eqa}{=^{+}}
\renewcommand{\lem}{<^{\ast}}
\newcommand{\gem}{>^{\ast}}
\newcommand{\eqm}{=^{\ast}}
\newcommand{\lel}{<^{\log}}
\newcommand{\gel}{>^{\log}}
\newcommand{\eql}{=^{\log}}

\newcommand{\BB}{\mathbf{BB}}
\newcommand\D{{\mathbf{D}}}
\newcommand\Q{\mathbb{Q}}
\newcommand\R{\mathbb{R}}
\newcommand\Z{\mathbb{Z}}
\newcommand\C{\mathbb{C}}
\newcommand\M{\mathbf{M}}
\newcommand\h{\mathbb{h}}
\newcommand\N{\mathbb{N}}
\newcommand\BT{\{0,1\}}
\newcommand\FS{\BT^*}
\newcommand\IS{\BT^\infty}
\newcommand\FIS{\BT^{*\infty}}

\newcommand\Hl{{\mathbf H}}
\newcommand\Hu{\overline{\mathbf H}}

\newcommand\ml{\underline{\mathbf m}}

\newcommand\uhr{\upharpoonright}
\renewcommand\H{\mathbf{H}}
\renewcommand\R{{\mathbb R}}
\newcommand\Hq{{\mathbf H}_\mathbf{q}}
\newcommand\Hd{{\mathbf H}_\mathbf{d}}
\newcommand\K{{\mathbf K}} 
\newcommand\QC{{\mathbf {QC}}}
\renewcommand\i{{\mathbf i}}
\newcommand\I{{\mathbf I}}
\newcommand\SI{{\mathbf i}}
\newcommand\m{{\mathbf m}}
\renewcommand\d{{\mathbf d}}
\renewcommand\h{{\mathbf h}}
\newcommand\ch{\mathcal{H}}
\newcommand\Ks{\mathbf{Ks}}

\author {Samuel Epstein\footnote{\href{mailto:samepst@jptheorygroup.org}{samepst@jptheorygroup.org}
}}
\title{\vspace*{-3pc} A Quantum EL Theorem} \date{\today}\maketitle
\begin{abstract}
In this paper, we prove a quantum version of the EL Theorem. It states that non-exotic projections of large rank must have simple quantum states in their images. A consequence to this is there is no way to communicate a quantum source with corresponding large enough von Neumann entropy without using simple quantum states.
\end{abstract}
\section{Introduction}

Quantum information theory studies the limits of communicating through quantum channels. In \cite{Holevo73}, the Holevo bound was proven, providing an upper bound on the amount of classical information shared between two parties that can prepare and measure mixed states. The Holevo bound states that only $n$ bits of classical information can be accessed from $n$ qubits. Schumacher's theorem \cite{Schumacher95} gives necessary and sufficient conditions under which there exists a reliable compression scheme to compress and decompress a quantum message with high fidelity. 

There is a large literature about the potential of quantum algorithms, with the most famous being Shor's factoring algorithm. There exists a relatively new area combining algorithms and quantum mechanics: the intersection of Algorithmic Information Theory (AIT) and Quantum Information Theory.  There are several interesting results in this new field. For example, in \cite{EpsteinQuantum21}, it was shown that given a quantum measurement (i.e. POVM) when it is applied to a pure quantum state, the vast majority of outcomes is meaningless random noise. 

This research program involves finding the quantum equivalent to definitions and theorems in AIT, with the primary concept being an quantum version of Kolmogorov complexity $\K(x)$. There are several such definitions that measure the algorithmic information content in a mixed or pure quantum state. In this paper we will use the definition $\K(\ket{\psi})$ in \cite{Vitanyi00}, which says a pure state $\ket{\psi}$ is complex if there is no simple (in terms of its classical enoding) pure state that has high quantum fidelity with $\ket{\psi}$. The results of this paper also applies to quantum algorithmic entropy, \cite{Gacs01}.
In \cite{Epstein19}, the quantum equivalent to algorithmic information and random deficiencies were defined. In addition conservation inequalities were proven with respect to unitary transform

In this paper we prove a Quantum EL Theorem. In AIT, the EL Theorem \cite{Levin16,EpsteinAP19} states that sets of strings that contain no simple member will have high mutual information with the halting sequence. It has many applications, including that all sampling methods produce outliers \cite{Epstein21}. The Quantim EL Theorem states that non exotic projections of large rank must have simple quantum pure states in their images. By non exotic, we mean the coding of the projection has low information with the halting sequence.

The Quantum EL Theorem can be used to address open issues in Quantum Information Theory. In \cite{Gacs01} the following remark was made.\\

\noindent\textbf{Remark.} (\cite{Gacs01}). \textit{Maybe the study of the problem for quantum description complexity helps with the understanding of the problem for von Neumann entropy, and its relation to coding tasks of quantum information theory.}\\

The theorem in this paper helps address this remark.\\

\noindent\textbf{Claim.} \textit{As the von Neumann entropy associated with the quantum source increases, the lossless quantum coding projectors have larger rank and thus must have simpler (in the algorithmic quantum complexity sense) pure states in their images.}

\section{Related Work}
For information about the history and foundation of algorithmic information theory, we refer readers to the textbooks \cite{DowneyHi10} and \cite{LiVi08}. There are several definitions that model the algorithmic content of a quantum state. In \cite{BerthiaumeVaLa01}, the complexity of a quantum state is equal to the size of the smallest quantum Turing machine that can approximate the state to a given fidelity. In \cite{MoraBe05}, the algorithmic complexity of a quantum state is equal to the minimal length of an encoding of the preparation of the state through quantum gates. In \cite{Gacs01}, the algorithmic entropy of a quantum state is measured by the negative logarithmic of the state multiplied by a universal lower computable semi-density matrix. In \cite{Vitanyi00}, the entropy of a pure quantum state is equal to the classical complexity of an elementary approximating state plus the negative logarithm of their fidelity. A quantum version of Brudno's theorem was proven in \cite{Benatti06}. Randomness for infinite quantum spin chains, called quantum Martin L\"{o}f random sequences, was introduced in \cite{NiesSc19}. An infinite version of algorithmic entropy can be found at \cite{Benatti14}. 

\section{Conventions}
The length of a string $x\in\FS$ is $\|x\|$. For positive real function $f$, $\lea f$, $\gea f$, and $\eqa f$ is used to represent $< f+O(1)$, $>f+O(1)$, and $=f\pm O(1)$. The encoding of $x\in\FS$ is $1^{\|x\|}0x$. For the nonnegative real function $f$, the terms ${\lel}f$, ${\gel} f$, and ${\eql}f$ represent the terms ${<}f{+}O(\log(f{+}1))$, ${>}f{-}O(\log(f{+}1))$, and ${=}f{\pm}O(\log(f{+}1))$, respectively. 

For strings $x,y\in\FS$, the output of algorithm $T$ on input $x$ and auxilliary input $y$ is denoted $T_y(x)$. An algorithm $T$ is prefix free iff for strings $x,y,s\in\FS$, $\neq\emptyset$, if $T_y(x)$ halts then $T_y(xs)$ does not halt. There exists a universal prefix free algorithm $U$, where for all prefix-free algorithms $T$, there exists a $t\in\FS$, where for all $x,y\in\FS$, $U_y(tx)=T(x)$. This $U$ is used to define Kolmogorov complexity, with $\K(x|y)=\min\{\|p\|:U_y(x)=p\}$. The universal probability of $x\in\FS$, conditional to $y\in\FS$, is $\m(x|y)=\sum\{2^{-\|p\|}:U_y(p)=x\}$. By the coding theorem, we have $-\log \m(x|y)\eqa\K(x|y)$. We use $\I(x;\ch)=\K(x)-\K(x|\ch)$ to be the amount of information that the halting sequence $\ch\in\IS$ has about $x\in\FS$.

We use $\mathcal{H}_n$ to denote a Hilbert space with $n$ dimensions, spanned by bases $\ket{\beta_1},\dots,\ket{\beta_n}$. A qubit is a unit vector in the Hilbert space $\mathcal{H}_2$, spanned by vectors $\ket{0}$, $\ket{1}$. To model $n$ qubits, we use a unit vector in $\mathcal{H}_{2^n}$, spanned by basis vectors $\ket{x}$, where $x$ is a string of size $n$. 

A pure quantum state $\ket{\psi}$ of length $n$ is a unit vector in $\mathcal{H}_{2^n}$. Its corresponding element in the dual space is denoted by $\bra{\phi}$. The conjugate transpose of a a matrix $A$ is $A^*$. $\Tr$ is used to denote the trace of a matrix. 
Projection matrices are Hermitian matrices with eigenvalues in $\{0,1\}$. 

Pure quantum states are elementary if their values are complex numbers with rational coefficients, and thus they can be represented with finite strings. Thus elementary quantum states $\ket{\phi}$ can be enncoded as strings, $\langle\ket{\phi}\rangle$ and assigned Kolmogorov complexities $\K(\ket{\phi})$ and algorithmic probabilities $\m(\ket{\phi})$. They are equal the complexity (and algorithmic probability) of the strings that encodes the states. More generally, a complex matrix $A$ is elementary if its entries are complex numbers with rational coefficients and can be encoded as $\langle A\rangle$, and has a Kolmogorov complexity $\K(A)$ and algorithmic probability $\m(A)$.

\section{Quantum Projections}
Simplicity is measured according to the  classical information content of a pure state. It is similar to the definition in \cite{Vitanyi00} except a classical Turing machine is used instead of a quantum Turing machine.

\begin{dff}[Complexity of a Quantum Pure State]$ $\\
For $n$ qubit state $\ket{\phi}$, $\H(\ket{\phi}) = \min\{\K(\ket{\psi})- \log |\braket{\phi|\psi}|^2:\ket{\psi}\textrm{ is an elementary pure state}\}$.
\end{dff}

A probability is elementary if it has finite support and all its values are rational. The deficiency of randomness of a string $x$ with respect to an elementary probability mesaure $Q$ is $\d(x|Q) = \floor{-\log Q(x)}-\K(x|\langle Q\rangle)$. The stochasticity of a string is $\Ks(x) = \min_Q\{\K(Q)+3\log\d(x|Q)\}$.
\begin{lmm}[\cite{Epstein21,Levin16}]
\label{lmm:ks}
$\Ks(x)\lel \I(x;\ch)$.
\end{lmm}

\begin{thr}[Quantum EL Theorem]
\label{thr}
Fix an $n$ qubit Hilbert space. Let $P$ be a elementary projection of rank $>2^m$. Then, relativized to $(n,m)$,
$\min_{\ket{\phi}\in\mathrm{Image}(P)}\H(\ket{\phi})\lel 3(n-m)+\I( \langle P\rangle;\ch)$.
\end{thr}
\begin{prf}
We assume $P$ has rank $2^m$. Let $Q$ be the elementary probability measure that realized the stochasticity, $\Ks(P)$, of an encoding of $P$. We can assume that every string in the support of $Q$ encodes a projection of rank $2^m$. We sample $N$ independent pure states according to the uniform distribution $\Lambda$ on the $n$ qubit space. For each pure state $\ket{\psi_i}$ and projection $R$ in the support of $Q$, the expected value of $\bra{\psi_i}R\ket{\psi_i}$ is 
$$
\int \bra{\psi_i}R\ket{\psi_i}d\Lambda =\Tr R\int\ket{\psi_i}\bra{\psi_i}d\Lambda = 2^{-n}\Tr RI =2^{m-n}.
$$
Let random variable $X_R=\frac{1}{N}\sum_{i=1}^N\bra{\psi_i}R\ket{\psi_i}$ be the average projection size of the random pure states onto the projection $R$. Since $\bra{\psi_i}R\ket{\psi_i}\in [0,1]$ with expectation $2^{m-n}$, by Hoeffding's inequality, 
$$
\Pr(X_R\leq 2^{m-n-1}) <\exp \left[-N2^{-2(m-n)-1}\right]
$$ 
Let $d=\d(P|Q)$. Thus if we set $N = d2^{2(m-n)+1}$, we can find $N$ elementary $n$ qubit states such that $Q(\{R:X_R\leq 2^{m-n-1}\})\leq \exp(-d)$, where $X_R$ is now a fixed value and not a random variable. Thus $X_P>2^{m-n-1}$ otherwise one can create a $Q$-expectation test, $t$, such that $t(R)=\exp d$. This is a contradiction because 
$$
1.44d \lea \log(P) \lea \d(P|Q) \lea d,
$$
for large enough $d$ which we can assume without loss of generality. Thus there exists $i$ such that $\bra{\psi_i}P\ket{\psi_i}\geq 2^{m-n-1}$. Thus $\ket{\phi}=P\ket{\psi_i}/\sqrt{\bra{\psi_i}P\ket{\psi_i}}$ is in the image of $P$ and $|\braket{\psi_i|\phi}|^2=\bra{\psi_i}P\ket{\psi_i}\geq 2^{m-n-1}$. The elementary state $\ket{\psi_i}$ has classical Kolmogorov complexity $\K(\ket{\psi_i})\lel \log N + \K(Q,d)\lel 2(m-n)+\Ks(P)$. Thus by Lemma \ref{lmm:ks},
\begin{align*}
&\min\{\H(\ket{\psi}):\ket{\psi}\in\mathrm{Image}(P)\}\\
&\leq \H(\ket{\phi})\\
&\lel \K(\ket{\psi_i})+|\braket{\psi_i|\phi}|^2\\
&\lel 3(n-m)+\Ks(P)\\
&\lel 3(n-m)+\I(P;\ch).
\end{align*}
\qed
\end{prf}
\subsection{Computable Projections}

Theorem \ref{thr} is in terms of elementary described projecctions and can be generalized to arbitrarily computable projections. For a matrix $M$, let $\|M\|=\max_{i,j}|M_{i,j}|$ be the max norm. A program $p\in\FS$ computes a projection $P$ of rank $\ell$ if it outputs a series of rank $\ell$ projections $\{P_i\}_{i=1}^\infty$ such that $\|P-P_i\|\leq 2^{-i}$. For computable projection operator $P$, $\I(P;\ch)=\min\{\K(p)-\K(p|\ch):p\textrm{ is a program that computes }P\}$. 
\begin{lmm}[\cite{EpsteinDerandom22}]
\label{lmm}
For partial computable $f$, $\I(f(a);\ch)\lea \I(a;\ch)+\K(f)$.
\end{lmm}
\begin{cor}
\label{cor}
Fix an $n$ qubit Hilbert space. Let $P$ be a computable projection of rank $>2^m$. Then, relativized to $(n,m)$,
$\min_{\ket{\phi}\in\mathrm{Image}(P)}\H(\ket{\phi})\lel 3(n-m)+\I(P;\ch)$.
\end{cor}
\begin{prf}
Let $p$ be a program that computes $P$. There is a simply defined algorithm $A$, that when given $p$, outputs $P_n$ such that $\min_{\ket{\psi}\in\mathrm{Image}(P)}\H(\ket{\psi})\eqa\min_{\ket{\psi}\in\mathrm{Image}(P_n)}\H(\ket{\psi})$. Thus by Lemma \ref{lmm}, one gets that $\I(P_n;\ch)\lea \I(P;\ch)$. The corollary follows from Theorem \ref{thr}.
\qed
\end{prf}
\section{Quantum Data Compression}
A quantum source consists of a set of pure quantum states $\{\ket{\psi_i}\}$ and their corresponding probabilities $\{p_i\}$, where $\sum_ip_i=1$. The pure states are not necessarily orthogonal. The sender, Alice wants to send the pure states to the receiver, Bob. Let $\rho = \sum_ip_i\ket{\psi_i}\bra{\psi_i}$ be the density matrix associated with the quantum source. Let $S(\rho)$ be the von Neumann entropy of $\rho$. By Schumacher compression, \cite{Schumacher95}, in the limit of $n\rightarrow\infty$, Alice can compress $n$ qubits into $S(\rho)n$ qubits and send these qubits to Bob with fidelity approaching 1. For example, if the message consists of $n$ photon polarization states, we can compress the inital qubits to $nS(\rho)$ photons. Alice cannot compress the initial qubits to $n(S(\rho)-\delta)$ qubits, as the fidelity will approach 0. The qubits are compressed by projecting the message onto a typical subspace of rank $nS(\rho)$ using a projector $P$. The projection occurs by using a quantum measurement consisting of $P$ and a second projector $(\mathbf{1}-P)$, which projects onto a garbage state. 
\begin{quote}
\textit{The results of this paper says that as $S(\rho)$ increases, there must be simple states in the range of $P$. There is no way to communicate a quantum source with large enough $S(\rho)$ without using simple quantum states.}
\end{quote}

\end{document}